\documentclass[prb,preprint,showpacs,superscriptaddress,amsmath,amssymb]{revtex4}

\usepackage{amsmath}

\usepackage{graphicx,amssymb} 
\usepackage{epsfig}
\begin{document}
\title[Atomistic study of the encapsulation of diamondoids inside carbon nanotubes]{Atomistic study of the encapsulation of diamondoids inside carbon nanotubes}

\author{K S Troche, V R Coluci, D S Galv\~ao}

\address{Instituto de F\'{\i}sica ``Gleb Wataghin'',
Universidade Estadual de Campinas, Unicamp 13083-970, Campinas,
S\~ao Paulo, Brazil}

\date{\today}

\begin{abstract}
The encapsulation of hydrogen-terminated nanosized diamond fragments, the so-called diamondoids into armchair single walled carbon nanotubes with diameters in the range of 1.0 up to 2.2 nm has been investigated using classical molecular dynamics simulations. Diameter dependent molecular ordered phases were found for the encapsulation of adamantane (C$_{10}$H$_{16}$), diamantane (C$_{14}$H$_{20}$), and dihydroxy diamantane (C$_{14}$H$_{20}$O$_{2}$). The same types of chiral ordered phases (double, triple, 4- and 5-stranded helices) observed for the encapsulation of C$_{60}$ molecules were also observed for diamondoids. On the other hand, some achiral phases comprising layered structures were not observed. Our results also indicate that the modification of diamantane through functionalization with hydroxyl groups can lead to an enhancement of the packing of molecules inside the nanotubes compared to non-functionalized compounds. Comparisons to hard-sphere models are also presented revealing differences, specially when more asymmetrical diamondoids are considered. For larger structures (adamantane tetramers) we have not observed long-range ordering for nanotubes with diameters in the range of 1.49 to 2.17 nm but only a tendency to form incomplete helical structures. 
\end{abstract}


\maketitle

\section{Introduction}
The discovery of the C$_{60}$ fullerene \cite{1}, carbon nanotubes \cite{2,2a} and the first realization of encapsulation of C$_{60}$ in carbon nanotubes \cite{3} have generated high interest in many research areas in nanotechnology. The large amount of experimental and theoretical works for these materials have been demonstrated their variety of functionalities making them a new and interesting class of nanostructured materials.

The structure of single walled carbon nanotubes (SWCNTs) containing a linear arrangement of C$_{60}$ fullerenes, generically called ``nano-peapods'' \cite{3}, has motivated experimental studies of ordered phases of C$_{60}$ inside boron nitride \cite{4} and carbon nanotubes \cite{10} of large diameters. A model of cylindrically confined hard spheres has been employed to demonstrate the emergence of ordered phase formation due to packing maximization \cite{6}. A systematic theoretical study using a continuum model was applied to the packing of C$_{60}$ inside SWCNTs and demonstrated that the formation of different ordered phases of C$_{60}$ molecules, resulting from the configuration of minimum energy of the system, is SWCNT diameter dependent \cite{7}. These phases consist of zigzag, double and triple helices, and more complex arrangements. Some of these predicted ordered phases have been experimentally observed by Khlobystov \textit{et al.} \cite{10} during the encapsulation of C$_{60}$ fullerenes inside double walled carbon nanotubes with internal diameters of from 1.1 up to 2.6 nm. Atomistic studies using molecular dynamics simulations of the encapsulation of C$_{60}$ \cite{8} and smaller fullerenes (C$_{20}$ and C$_{28}$) \cite{9} into SWCNTs have shown a good agreement with the hard-sphere and continuum models. In addition, atomistic studies of the encapsulation of C$_{70}$ and C$_{78}$ also showed the dependence on the tube diameter, revealing similar types of ordered arrangements as the ones observed in C$_{20}$, C$_{28}$, and C$_{60}$ cases, despite of the asymmetrical shapes of C$_{70}$ and C$_{78}$ molecules \cite{8}.

Recently, other carbon nanostructures, known as diamondoids, have also attracted the attention of the scientific community \cite{12,13}. Diamondoids are essentially hydrogen-terminated nanosized diamond fragments (Fig. 1). Adamantane (Fig. 1(a)), diamantane (Fig. 1(b)), and triamantane, present in petroleum crude oil at low concentrations \cite{13}, are the smaller diamondoids molecules, each one having only one isomer. The first structural elucidation of the smallest diamondoid, adamantane, occurred in 1913 \cite{15a}. It was found and isolated from Czechoslovakian petroleum in 1933 \cite{15c,15d} but only in 1957 it was synthesized \cite{15}. Small (1-2 nm in diameter) and large ($>$ 2 nm in diameter) diamondoids are found in multiples of adamantane. Larger diamondoids can appear in many shapes - rods, disks and even screws.  Adamantane oligomers \cite{17} and polymers present high stability, strength, and hardness. Due to their flexible chemistry, pure diamondoids are particularly attractive for preparation of pharmacological active compounds with pronounced membrane activity due to lipophilicity of the aliphatic diamondoid core structure \cite{13,18}.  On the other hand, functionalized diamondoids \cite{19} are ideal candidates for molecular building blocks in nanotechnology \cite{20}, and for molecular electronic applications when the incorporation of functionalized groups allows their attachment to metal surfaces to form self-assembled monolayers \cite{21}. 

McIntosh \textit{et al.} \cite{24} have studied the structural, electronic, and encapsulation properties of adamantane and diamantane using density functional theory. The encapsulation was investigated in ($n$,$n$) armchair SWCNTs ranging from $n=$ 4 up to 8. They found that the encapsulation of diamantane into armchair SWCNTs with diameter larger than 9.5 {\AA} ($n\geq$ 7) may occur spontaneously without any energetic cost. As pointed out by McIntosh \textit{et al.} \cite{24}, carbon nanotubes are promising structures to be used in assembling diamondoids. However, investigations of the diamondoid encapsulation in larger diameter SWCNTs ($n>8$) have not been yet carried out. For this reason, in this work, we have carried out atomistic molecular dynamics simulations to study the encapsulation of small diamondoids inside armchair SWCNTs to investigate the possibility of formation of ordered phases. 

A good agreement between the ordered phases predicted by the hard-sphere and continuum models, and atomistic simulations has been observed for the C$_{60}$ molecule due its spherical symmetry \cite{8}. On the other hand, one may not expect such agreement for asymmetric diamondoid structures. Aspects such as functionalization of diamondoids either by atom substitution or insertion, can lead to asymmetrical characteristics in geometrically symmetrical unmodified diamondoids. The asymmetry is more difficult to be addressed with analytical/continuum methods but can be easily and directly addressed in atomistic simulations.

\section{Methodology}
We performed classical molecular dynamics simulations using the universal force field \cite{25,26} implemented in the Cerius$^2$ package \cite{27} which includes van der Waals, bond stretch, bond angle bend, inversion
and torsional rotation terms. This methodology has been proved very effective in the dynamical aspects of carbon based nanostructures \cite{prl,oscnano,scroll}. The dynamics of diamondoids molecules within a cylindrically confined environment produced by the SWCNT is mainly due to van der Waals interactions. A Lennard-Jones 6-12 potential was used to describe the van der Waals interaction between carbon, hydrogen and oxygen atoms, i.e.,
\begin{equation}
\displaystyle E_{ij}^{vdW}(x)=D_{ij}\left[-2\left(\frac{x_{ij}}{x}\right)^6 +\left(\frac{x_{ij}}{x}\right)^{12}\right]
\end{equation}
where $x$ is the separation distance between the atoms $i$ and $j$, and 
\begin{equation}
\displaystyle D_{ij}=\sqrt{D_iD_j}\;\;\;\;\;\; x_{ij}=(1/2)(x_i+x_j),
\end{equation}
with $i$,$j$ = C (carbon), H (hydrogen), or O (oxygen). The following parameter values were used \cite{25,26}: $D_C=$ 0.105 kcal/mol, $x_C=$ 3.851 {\AA}, $D_H=$ 0.044 kcal/mol,  $x_H=$ 2.886 {\AA}, $D_O=$ 0.060 kcal/mol, and $x_O=$ 3.5 {\AA}.

We have considered ($n$,$n$) armchair SWCNTs with 8 $\leq$ $n$ $\leq$ 16, and length of 12 to 15 nm. Based on results for the C$_{60}$ \cite{8}, where the nanotube chirality presents a minor effect on the types of ordered phases, we also expect the same behavior for the encapsulation of adamantane and diamantane. The tubes were kept fixed during the simulations. We used the (8,8) nanotube as the smallest nanotube since the entry process of diamantane is energetically favorable for this nanotube \cite{24}. We have investigated the encapsulation of adamantane $T_d$ (C$_{10}$H$_{16}$, Fig. 1(a)), diamantane $C_{2h}$ (C$_{14}$H$_{20}$, Fig. 1(b)), functionalized dihydroxy diamantane (C$_{14}$H$_{20}$O$_{2}$, Fig. 1 (c)) \cite{28}, and 1,3-adamantane tetramer (C$_{40}$H$_{58}$, Fig. 1 (d)) \cite{17}.

For SWCNTs with $n<$ 12 we started the simulation by introducing the diamondoids one by one, until the complete filling of the nanotube. For the other cases, two or three linear arrangements of diamondoids have been initially inserted. The molecular dynamics simulations were then carried out in the canonical ensemble during hundreds of ps (time step=0.5 fs) at 300 K. For controlling temperature the Nos\'e thermostat \cite{nose} was used. To avoid local minima, the temperature has been gradually increased until 700 K by steps of 100 K per 50 ps. After that, a temperature decreasing of the system is carried out back to 300 K. At this final stage, all structures were geometrically optimized with convergence tolerances for energy and force of 10$^{-4}$ kcal/mol and 5$\times$10$^{-3}$ kcal/mol/{\AA}, respectively. The resulting optimized structures were used to identify the ordered phases.  

\begin{figure}
\begin{center}
\includegraphics[angle=0,width=70 mm]{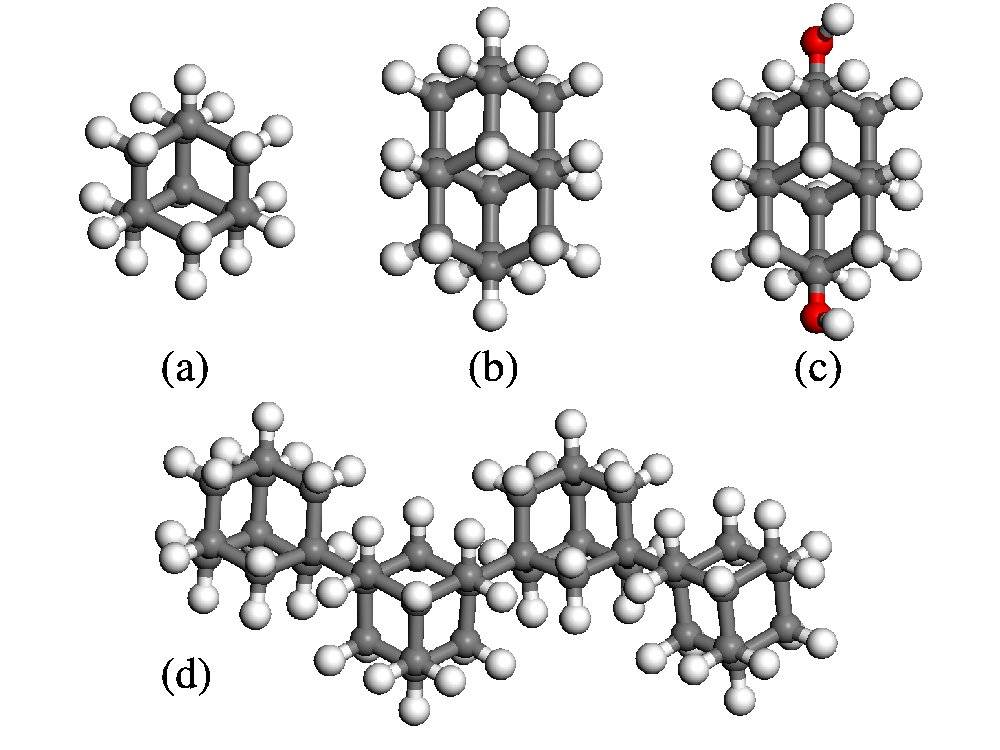}
\caption{Molecular structure of (a) adamantane, (b) diamantane, (c) dihydroxy diamantane, and (d) 1,3-adamantane tetramer. Carbon atoms are represented in gray, hydrogen in white, and oxygen in red.}
\end{center}
\end{figure}

\section{Results}
Adamantane and diamantane inside armchair SWCNTs with $n$ = 8 and $n$ = 9 were predicted to form a linear arrangements. Different ordered phases were obtained when $n>$ 9 to $n$ = 16. The obtained phases for these diamondoids molecules are listed in Table 1. We can see that the same type of ordered phases observed for C$_{60}$ molecules can also be present in the encapsulation of adamantane and diamantane, i.e, zigzag, double, triple, 4- and 5-stranded helices. However, some C$_{60}$ achiral phases such as two- and three-molecule layers were not observed for adamantane and diamantane for the SWCNTs considered here. For the C$_{60}$ encapsulation \cite{7}, these phases appear between double and triple helix phases, and triple and 4-stranded helix phases, respectively. For adamantane and diamantane, we can see a direct transition from double to triple helix phase and from triple to 4-stranded helix phase without these intermediate achiral phases. Excluding the linear and zigzag achiral phases, our simulation results suggest that adamantane and diamantane show a tendency to preferentially form chiral phases, at least until diameters up to about 2.2 nm. It is an indication that not only the nanotube diameter but also the structure of the encapsulated molecule play important roles on the ordered phase type that eventually is formed within the SWCNT.

Figures 2 and 3 show some of the ordered phases obtained for adamantane and diamantane, respectively. For the (16,16) tube (Fig. 2), the adamantane molecules form a cylindrical shell consisting of a 5-stranded helix on the internal nanotube wall. This phase has been predicted by Hodak and Girifalco as the minimum energy configuration of C$_{60}$ fullerenes inside (21,21) SWCNT \cite{7}. It is important to notice the presence of several holes (Figs. 2 and 3), corresponding to missing molecules, in the packing configurations. These holes along helix configurations appear to be most frequently observed for diamantane arrangements. Despite the presence of such defects, which cause a local disorder in the arrangement, it is still possible to observe some degree of organization that indicates the existence of an ordered phase. From the geometrical point of view, an analogy between adamantane and C$_{60}$, and between diamantane and C$_{70}$ can be made due to the more pronounced asymmetry presented by diamantane and C$_{70}$ molecules. In fact, more disorder is observed in the arrangement presented by the encapsulation of diamantane than in the case of adamantante, similarly to the cases of ordered phases of C$_{60}$ and C$_{70}$ \cite{8}. This disorder plays an important role in the phase formation as seen for the specific case of (16,16) SWCNT encapsulated with diamantane (Figure 3). In this case, an incomplete 4-stranded helix is observed, acting as a intermediate phase between a 4- and 5-stranded helix, the former appearing in the (15,15) and the latter expected to be possibly formed in the (17,17) SWCNT. 

\begin{table}[ht]
\caption{Packing phases obtained from molecular dynamics simulations of diamondoids inside armchair SWCNTs of different diameters. The arrangements presented by C$_{60}$ molecules are also shown for comparison \cite{8}.}
\vspace{0.05cm} \centering
\label{t01} \footnotesize
\begin{tabular}{c c c c c}
 \hline
Nanotube  &   C$_{60}$ fullerene  & Adamantane & Diamantane & Dihydroxy-diamantane \\
\hline
$(8,8)$		&$-$  & Linear & Linear  & Linear\\
$(9,9)$		&$-$  & Linear & Linear  & Linear\\
$(10,10)$	& Linear &  Zigzag& Zigzag  & Linear \\
$(11,11)$	& Zigzag & Zigzag & Zigzag & Zigzag \\
$(12,12)$	& Zigzag &  Double helix & Zigzag & Double helix \\
$(13,13)$	& Zigzag &  Triple helix &  Double helix & Double helix\\
$(14,14)$	& Zigzag & 4$-$stranded helix & Triple helix & Triple helix\\
$(15,15)$	& Zigzag & 5$-$stranded helix & 4$-$stranded helix & 4$-$stranded helix\\
$(16,16)$	& Double helix &  5$-$stranded helix& Incomplete 4$-$stranded helix & 5$-$stranded helix\\
\hline
\end{tabular}
\end{table}

\begin{figure}
\begin{center}
\includegraphics[angle=0,width=90 mm]{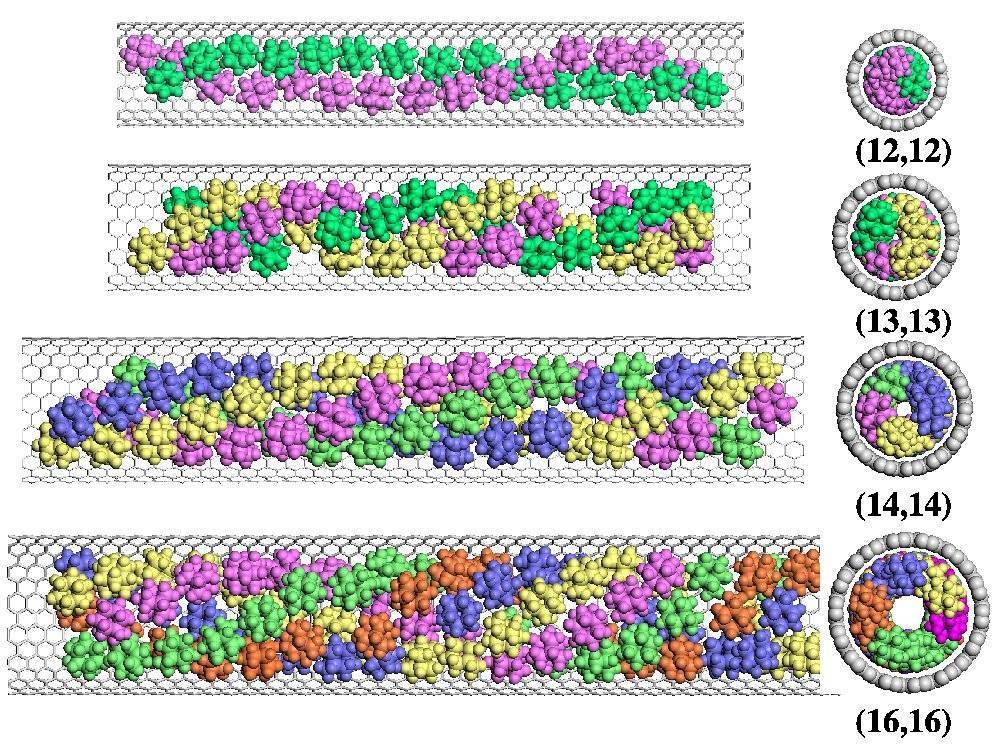}
\caption{Double, triple, 4-stranded, and 5-stranded helix molecular phases obtained from molecular dynamics simulations of adamantane encapsulated in (12,12), (13,13), (14,14), and (16,16) SWCNTs, respectively. Different colors are used to represent the helical structures.}
\end{center}
\end{figure}

\begin{figure}
\begin{center}
\includegraphics[angle=0,width=90 mm]{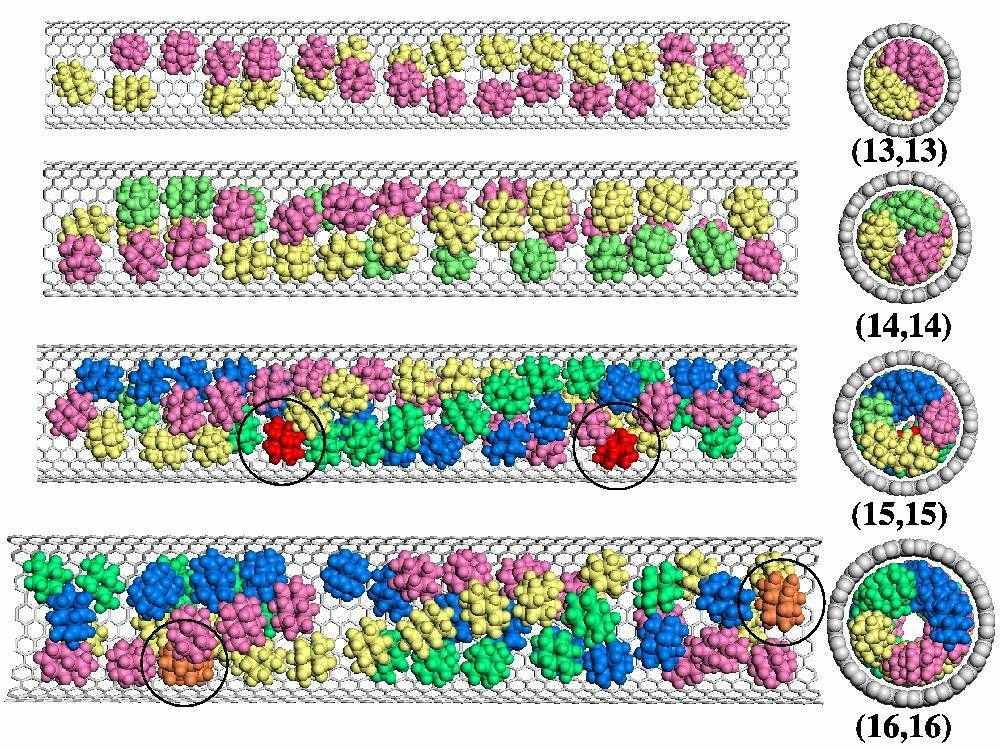}
\caption{Double, triple, 4-stranded and incomplete 4-stranded helix molecular phases obtained from molecular dynamics simulations of diamantane encapsulated in (13,13), (14,14), (15,15), and (16,16) SWCNTs, respectively. Circles indicate molecules that were found to not belong to any of the identified helices of the arrangement.}
\end{center}
\end{figure}

In addition to pharmacological applications of functionalized diamondoids, McIntosh \textit{et al.} have proposed that they can be used to increase the interaction between adjacent diamondoids within a SWCNT \cite{24}. The functionalization would enhance bonding and would allow the connection between encapsulated diamondoids, favoring a diamondoid polymerization inside the encapsulating tube. They proposed the substitution of terminal carbon atoms by boron and nitrogen in adjacent diamantanes in a linear arrangement. For this case, they obtained a value of 1.64 eV for the binding energy of functionalized diamantanes in contrast to 0.13 eV of the unmodified ones. While this type of functionalization has not yet been produced, other types have been already synthesized. An example is the dihydroxy diamantane (Fig. 1 (c)) \cite{28}. Preliminaries calculations involving the reaction of two dihydroxy diamantane molecules to form the C$_{14}$H$_{18}$O dimer through the release of water have indicated a highly endothermic process, which would forbide the diamantane polymerization. In despite of that, we have investigated the encapsulation properties of the dihydroxy diamantane since hydroxyl group usually enhances the molecular reactivity that could allow diamondoid polymerization inside the nanotube through alternative reaction pathways.

We have also observed the formation of ordered phases for the dihydroxy diamantane but with some differences with respect to the unmodified diamantane. The obtained phases are also shown in Table 1 and some of them presented in Figure 4. The dihydroxy diamantane forms a linear structure inside armchair nanotubes with 8 $\leq n \leq$ 10. For $n\geq$ 9 local disorder was observed to be more pronounced. A zigzag arrangement was observed when the molecules are confined into (11,11) SWCNT. The double helix arrangement was obtained for (12,12) SWCNTs. For the same SWCNT a zigzag pattern was found instead for the unmodified diamantane. We can also see a double helix arrangement in (13,13) SWCNT encapsulated with dihydroxy diamantane but with a different pitch of the one of the helices observed in (12,12) SWCNT. However, a similar pitch is found for the helices presented in (13,13) SWCNT encapsulated with dihydroxy diamantane and diamantane, respectively. Increasing even further the SWCNT diameter, complex helical arrangements are also formed such as a 5-stranded helix in (16,16) SWCNT, instead of the 4-stranded helix observed for unmodified diamantane molecules encapsulated into the same SWCNT. From these results we can conclude that the functionalization with hydroxyl groups in diamantane (Fig. 1 (c)) seems to optimize the packing.

\begin{figure}
\begin{center}
\includegraphics[angle=0,width=90 mm]{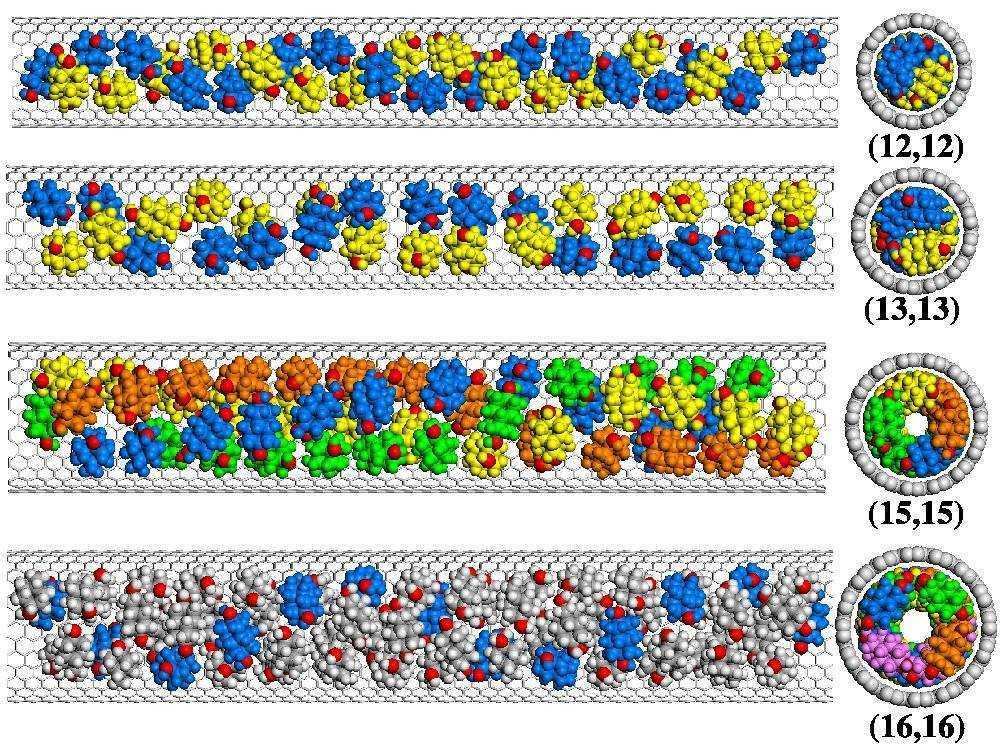}
\caption{Double, 4-stranded and 5-stranded helix molecular phases obtained from molecular dynamics simulations of dihydroxy diamantane encapsulated in (12,12) and (13,13), (15,15), and (16,16) SWCNTs, respectively.} 
\end{center}
\end{figure}

For the more asymmetrical diamondoid-based compound investigated here, i.e., 1,3-adamantane tetramer (Fig. 1 (d)) we have found that the smallest tube where it can be encapsulated is the (10,10) (Fig. 5). We have observed the formation of cylindrical shells around of the SWCNT internal wall with some tendency to form helices for (11,11) up to (16,16) SWCNTs. However, in general, we could not see the presence of totally complete ordered phases during the simulation time used in our simulations. For instance, in Figure 5, we can see an incomplete 4-stranded helix arrangement in (16,16) SWCNT where the incomplete helix (yellow) appears due to some misalignment of some molecules inside the nanotube (inside the circle).  

\begin{figure}
\begin{center}
\includegraphics[angle=0,width=90 mm]{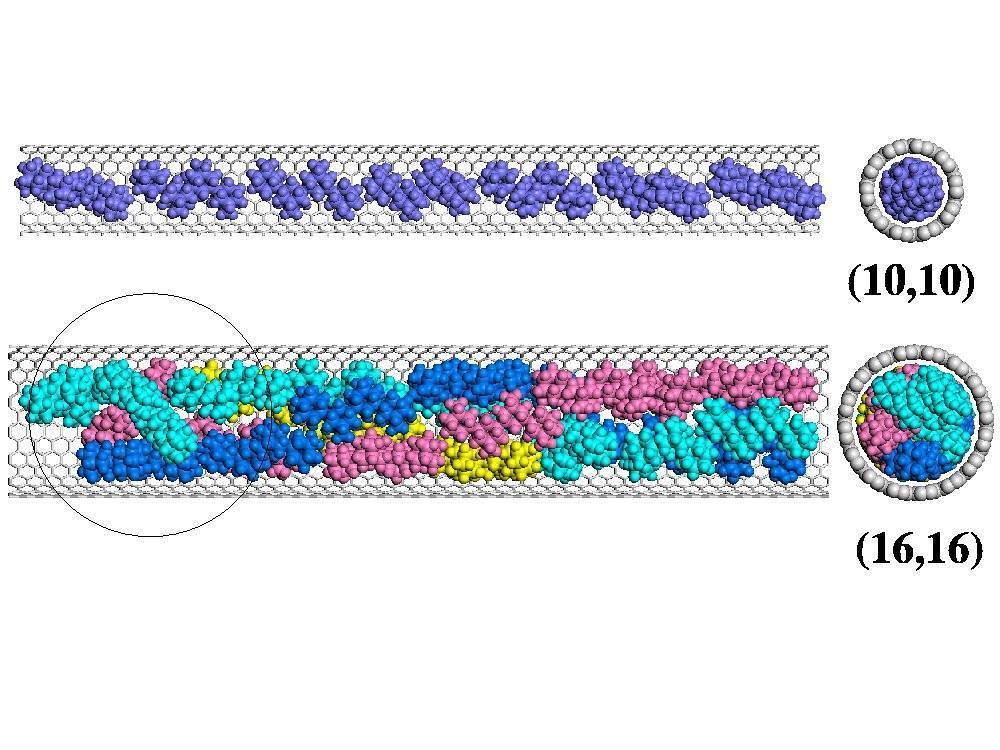}
\caption{Linear chain and incomplete 4-stranded helix molecular phases obtained from molecular dynamics simulations of 1,3-adamantane tetramer encapsulated in (10,10) and (16,16) SWCNTs, respectively. The circle indicates molecules that are misaligned with the helix pattern.} 
\end{center}
\end{figure}

Analyzing the volume fraction \cite{vol} as a function of nanotube diameter was possible to compare the C$_{60}$ packing with results obtained from the hard-sphere model \cite{8}. A good agreement with the hard-sphere model was obtained for SWCNTs with $n\geq$ 15, when the ordered phases are three-dimensional \cite{8}. In order to make comparisons between the packing of diamondoids predicted by atomistic simulations and by the hard-sphere model, we have used the value of the available SWCNT diameter $D = D_n - \Delta$, where $D_n$ is the diameter of the SWCNT optimized with the universal force field \cite{8}, and $\Delta$ = 3.38 {\AA} is the thickness of the nanotube wall. For the hard-sphere model $D$ is written in units of the ball diameter \cite{6}. For each diamondoid curve we have expressed $D$ in units of the its molecular diameter $d$ assuming a spherical shape, i.e., $d=2(3V/4\pi)^{1/3}$, where $V$ is the molecular volume comprised by the Connolly surface \cite{volume}. The obtained volume values for adamantane, diamantane, dihydroxy diamantane, and C$_{60}$ were (in {\AA}$^3$) 156.8, 205.1, 217.1, and 490.6, respectively. With the volume calculated for the C$_{60}$ molecule by this method, the corresponding value of the C$_{60}$ diameter is in agreement with the intermolecular distance 0.97 $\pm$ 0.02 nm obtained for the linear arrangement in the (10,10) SWCNT \cite{8}.

Figure 6 shows the volume fraction as function of $D$ for the encapsulation of adamantane, diamantane, dihydroxy diamantane, and C$_{60}$ molecules. We can see that all diamondoids for the linear arrangement in the (8,8) SWCNT ($D=$ 1) present a volume fraction smaller than $2/3$ (the expected value for the encapsulation of hard spheres in a hard cylinder of the same sphere diameter). Due to the spherical symmetry presented by the C$_{60}$ molecule, a reasonable agreement with the hard-sphere model is seen for $D >$ 1.7 . For $D <$ 1.7 an increase in the volume fraction is observed for the C$_{60}$ packing. In this region the phases are basically two-dimensional and, therefore, the number of encapsulated fullerenes is limited by the nanotube length. As a direct consequence, increasing the nanotube diameter causes the volume fraction value to decrease. The behavior of the volume fraction shown by the adamantane encapsulation is qualitatively similar to the C$_{60}$, which can be associated with an approximated spherical shape also shown by the adamantane. In both cases we can see an increase of the volume fraction during the transition of the zigzag to double helix arrangements ($D\simeq 1.73$). Furthermore, the behavior for the adamantane is qualitatively different for the other diamondoids considered here. For both diamantane and dihydroxy diamantane, the volume fraction begins to increase after $D\simeq 1.93$ where the double helix arrangement is present ((13,13) SWCNT). The effect of functionalization also reflects on the volume fraction values. For the functionalized diamantane (dihydroxy diamantane) we can see higher values of the volume fraction than the unmodified diamantane, indicating a packing enhancement produced by the introduction of hydroxyl groups in diamantane. This enhancement is more evident for the achiral linear and zigzag phases ($D\lesssim 1.7$).

\begin{figure}
\begin{center}
\includegraphics[angle=0,width=90 mm]{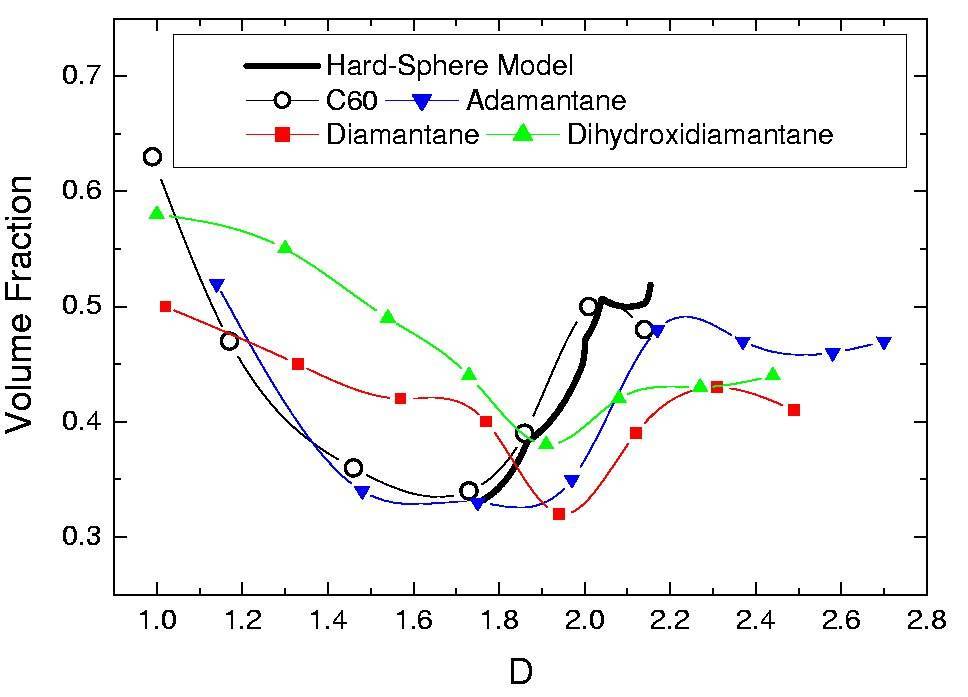}
\caption{Volume fraction as function of the available diameter $D$ expressed in units of each molecular diameter. The lines connecting the points are only to help visualization.} 
\end{center}
\end{figure}

\section{Conclusions}
Using atomistic molecular dynamics simulations we were able to identify molecular ordered phases on the encapsulation of diamondoids into armchair single walled carbon nanotubes. We have verified that the arrangement types depend on the diameter of the encapsulating nanotube as well as the structure of the encapsulated molecule. For the SWCNTs considered in this work the investigated diamondoids have not shown some achiral phases that have been previously predicted on the encapsulation of C$_{60}$ molecules. The asymmetry presented by the diamondoids can lead to local disorders on the arrangement that do not affect the long-range ordering. An enhancement on the packing and a favoring to chiral phases were obtained in the encapsulation of functionalized diamondoids. The more packed phases shown by functionalized diamondoids may serve as candidates for diamondoid polymerization inside carbon nanotubes. The encapsulation of larger structures (1,3-adamantane tetramers) was only found  for tubes with diameters larger than 1.36 nm ((10,10) SWCNT). For those diamondoid-based polymers we have not observed long-range ordering for diameters larger than 1.49 nm but only a tendency to form incomplete helical structures. This long-range ordering is precluded by misaligned molecules within the helical arrangements.

We acknowledge the financial support from the Instituto de
Nanosci\^encias, THEO-NANO, IMMP/CNPq, the Brazilian agencies FAPESP, Capes, and CNPq.


\begin{thebibliography}{28}
\bibitem{vol}The volume fraction is ratio of the volume of $N$ packed molecules ($NV$) and the nanotube volume, i.e., $NV/(\pi D^2L/4)$, where $L$ is the nanotube length and $D$ is the nanotube diameter.
\bibitem{volume} The Connolly surface is at the boundary between a probe and the atoms as the probe rolls over the surface that intersects with the van der Waals radii of the atoms in the structure. The calculations were performed in Cerius$^2$\cite{27} using a probe of radius of 1.0 {\AA} with a grid resolution of 0.15 {\AA}.
\end{thebibliography}
\end{document}